\documentclass[fleqn,10pt]{wlscirep}
\usepackage[utf8]{inputenc}
\usepackage[T1]{fontenc}
\usepackage{lineno}
\usepackage{amsmath}
\usepackage{subcaption}
\usepackage{graphicx}

\usepackage[format=plain,justification=justified,singlelinecheck=false,font={stretch=1.125,small,sf},labelfont=bf,labelsep=space]{caption}

\def\etal{{\it et al. }}
\newcommand{\ie}[0]{\textit{i.e.}, }

\newcommand{\eg}[0]{\textit{e.g.}, }

\newcommand{\via}[0]{\textit{via} }

\usepackage{chemformula} 
\usepackage{braket}
\newcommand{\DefineAuthor}[2]{%
  \expandafter\newcommand\csname #1note\endcsname[1]{%
    \textbf{\textcolor{#2}{\textbf{#1:} ##1}}}%
  \expandafter\newcommand\csname #1\endcsname[1]{
    \textbf{\textcolor{#2}{##1}}}
  \expandafter\newcommand\csname #1cancel\endcsname[1]{%
    \textbf{\textcolor{#2}{\sout{##1}}}}%
  \expandafter\newcommand\csname #1change\endcsname[2]{%
    \textbf{\textcolor{#2}{\sout{##1} ##2}}}%
  \newenvironment{#1text}{\color{#2}}{\color{black}}
}
\definecolor{dartmouthgreen}{rgb}{0.05, 0.5, 0.06}
\DefineAuthor{LMS}{red}
\DefineAuthor{AF}{brown}
\DefineAuthor{AS}{blue}

\usepackage{xcolor}

\usepackage{algpseudocode}%
\usepackage{listings}%
\usepackage{siunitx}%
\DeclareSIUnit\electronvolt{eV}
\DeclareSIUnit\meva{\milli \electronvolt \per \angstrom}
\usepackage{textgreek}%
\usepackage{multirow}

\title{Data-Driven Thermal and Mechanical Modeling of Defective Covalent Organic Frameworks}

\author[1]{Aleksander Szewczyk}
\author[1,*]{Leonardo Medrano Sandonas}
\author[1]{David Bodesheim}
\author[2,3]{Bohayra Mortazavi}
\author[1,4,5,6,*]{Gianaurelio Cuniberti}
\affil[1]{Institute for Materials Science and Max Bergmann Center of Biomaterials, TUD Dresden University of Technology, 01062, Dresden, Germany}
\affil[2]{Institute of Photonics, Department of Mathematics and Physics, Leibniz Universität Hannover, Welfengarten 1A, 30167, Hannover, Germany}
\affil[3]{Cluster of Excellence PhoenixD, Leibniz Universität Hannover, Welfengarten 1A, Hannover 30167, Germany}
\affil[4]{Dresden Center for Computational Materials Science (DCMS), TUD Dresden University of Technology, 01062 Dresden, Germany}
\affil[5]{Cluster of Excellence CARE, TU Dresden and RWTH Aachen, Germany}
\affil[6]{Cluster of Excellence CeTI, TU Dresden, Germany}
\affil[*]{Corresponding authors:  Leonardo Medrano Sandonas (leonardo.medrano@tu-dresden.de), Gianaurelio Cuniberti (gianaurelio.cuniberti@tu-dresden.de)}

\begin{abstract} 
Covalent Organic Frameworks (COFs) are versatile two-dimensional (2D) materials for flexible electronics, catalysis, and sensing, owing to their tunable architectures and large surface areas. However, like most materials, COFs inevitably contain synthesis-induced defects, which--similar to graphene--can strongly influence intrinsic properties, such as thermal transport and mechanical strength.
To address this challenge, we have assessed the performance of a set of machine learning interatomic potentials (MLIP) capable of efficient large-scale simulations of COFs with quantum accuracy.
In doing so, QCOF models ("Quantum COF") were developed by tuning the state-of-the-art MACE architecture on an extensive dataset of non-equilibrium COF conformations generated from high-fidelity density functional theory calculations.
The accuracy, computational efficiency, memory footprint, and transferability to unseen chemical environments of these models were benchmarked against general-purpose MACE models (MACE-OFF24 and MACE-MPA-0) and their fine-tuned variants. 
Our results show that an invariant QCOF model with a small descriptor dimensionality and cutoff outperforms all other models in most validation tasks, including scalability to large systems, force prediction in defective COFs, and phonon dispersion calculations.
The best-performing QCOF model was then used to run large-scale simulations of thermal conductivity for defective CTF-1 and COF-LZU1 systems (>40k atoms) \via non-equilibrium MD, revealing a more pronounced sensitivity of CTF-1 to structural defects.
Stress–strain curves were also investigated, showing that the mechanical response remains nearly invariant at low defect densities, while asymmetric behaviour emerges at large strains.
This work thus provides a foundation for the design of robust and high-performance quantum-informed MLIP for large-scale property simulations of pristine and defective of extended network materials.

\end{abstract}

\begin{document}

\flushbottom
\maketitle

\thispagestyle{empty}

\section*{Introduction} \label{sec:intro} 

\begin{figure}[ht!]
\centering
\includegraphics{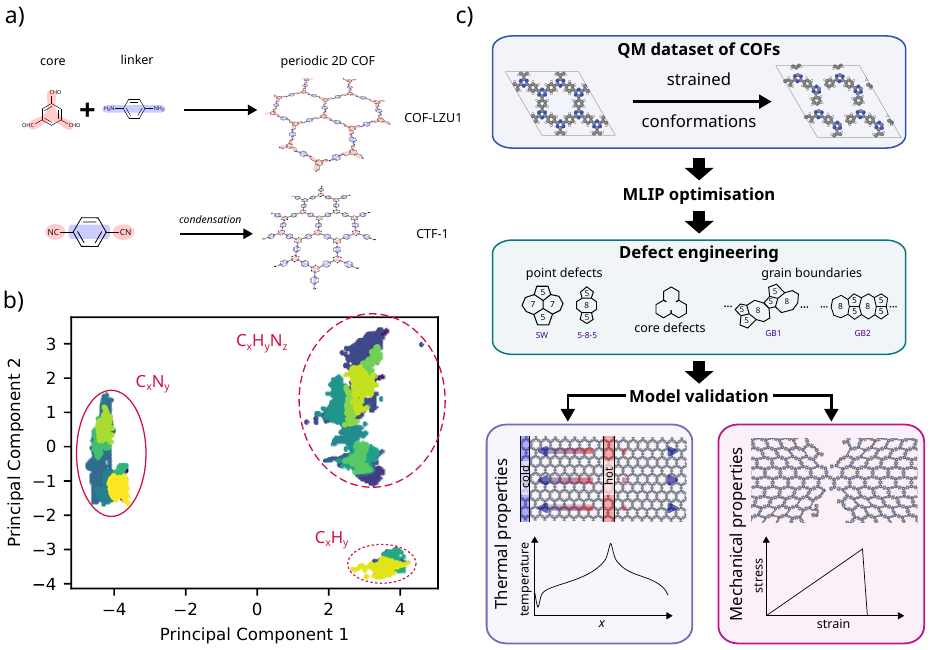}
\caption{a) Covalent organic frameworks (COFs) consist of cores and/or linkers arranged in specific geometries. Their properties can be readily tuned by modifying these secondary building units, as illustrated here for COF-LZU1 and CTF-1. b) Principal component analysis (PCA) of the 36,000 COF conformations in the training dataset. Compounds with different chemical compositions form distinct clusters, as indicated by the color coding (23 carbon nitride nanosheets). c) Schematic of the computational workflow used in this work, from the quantum-mechanical database, \via training of a machine learning interatomic potential (MLIP), to the prediction of thermal and mechanical properties.}
\label{fig:01-intro-to-cofs}
\end{figure}

Covalent Organic Frameworks (COFs) are a class of porous materials characterized by high surface area, tunable structural properties, and well-defined porosity. Their fully organic composition also makes them attractive from a sustainability perspective. These features have led to a wide range of applications, including gas storage \cite{furukawaStorageHydrogenMethane2009}, flexible electronics \cite{wuRecentProgressCovalent2024}, membrane-based separations \cite{yuanCovalentOrganicFrameworks2019}, and catalysis \cite{guoCovalentOrganicFrameworks2020}.
COFs are constructed from organic building blocks that assemble into periodic networks through strong covalent bonds (see Fig. \ref{fig:01-intro-to-cofs}). Depending on their topology, they can form either three-dimensional frameworks or two-dimensional layered structures, in which individual sheets are held together by van der Waals interactions. In this work, we focus on single-layer COFs, which offer distinct advantages over their three-dimensional counterparts. In particular, the anisotropy of bonding enables exfoliation into ultrathin films with nanometer-scale thickness, thereby maximizing the exposure of active sites.

The properties of COFs can be systematically tuned through the choice of their molecular building blocks, resulting in a vast design space.
This tunability makes computational modelling an essential tool for exploring structure-property relationships and guiding material design. However, the large, complex, and often flexible unit cells of COFs pose significant challenges for atomistic simulations.
A range of computational approaches has been applied to their study, including Density Functional Theory (DFT) \cite{niEngineeringFlatBands2022, mourinoSearchCovalentOrganic2023, heidariDFTStudyCOF12023}, tight-binding methods \cite{lukoseReticularConstructionConcept2010, raptakisPredictingBulkModulus2021, bodesheimHierarchiesHofstadterButterflies2023, bittner_engineering_2024}, classical force fields \cite{ghahariProposingTwodimensionalCovalent2022, raptakisPredictingBulkModulus2021}, and coarse-grained models \cite{durholtCoarseGrainingForce2016, mohamedCoarsegrainedForceField2024, alvaresCoarsegrainedModelingZeolitic2023, bodesheimElasticPropertiesDefective2025, cranfordExtendedGraphynesSimple2012, raptakisPredictingBulkModulus2021, croyCoarseGrainedElasticitySingleLayer2022}.
First-principles methods such as DFT provide accurate electronic structure information but are computationally demanding and scale poorly with system size, typically limiting simulations to a few hundred atoms. In contrast, more efficient approaches, such as coarse-grained models, enable simulations at larger length scales and can capture phenomena that are inaccessible to atomistic methods, including structural disorder and defects.
Experimentally, COFs frequently exhibit crystal imperfections such as missing cores or linkers, Stone–Wales-type defects, grain boundaries, and irregular layer stacking \cite{daliranProbingDefectsCovalent2024, prakashReviewCovalentOrganic2024}. As a result, idealized periodic models often fail to accurately represent real materials.
These defects can strongly influence mechanical properties as well as transport processes governed by the scattering of phonons \cite{scottPhononScatteringEffects2018, polancoInitioPhononPoint2018}, excitons, and charge carriers \cite{sorkinEffectsPointDefect2022, liuEffectDefectsOptical2023, kunstmannLocalizedDefectStates2017}. Accurately capturing such effects therefore requires simulations of systems containing up to tens of thousands of atoms, well beyond the practical limits of DFT.

For calculating electronic properties, efficient alternatives to first-principles methods are provided by semi-empirical approaches such as density-functional tight-binding (DFTB) \cite{duImpactStructuralDefects2024, raptakisPredictingBulkModulus2021}.
In contrast, non-electronic properties, including thermal transport and elastic properties, are typically studied using molecular mechanics force fields, which do not account for electronic interactions in their design \cite{thakolkaranDeepLearningReveals2025, kwonThermalConductivityCovalentOrganic2023, mortazaviFirstPrinciplesMultiscaleModeling2021, mortazaviMachinelearningInteratomicPotentials2020}.
Conventionally, force fields employ simple, fixed functional forms and are parameterized using a limited set of reference systems. They remain widely used, particularly in biomolecular simulations, where access to long timescales (beyond nanoseconds) and large system sizes (up to millions of atoms) is essential \cite{poma25}.
However, their inherent rigidity limits their ability to accurately capture complex interatomic interactions, resulting in reduced accuracy for conformational energies and atomic forces. Moreover, many commonly used force fields, such as the Universal Force Field \cite{rappeUFFFullPeriodic1992}, do not describe adequately bond breaking and formation, which is crucial for investigating mechanical strength.
This limitation highlights a fundamental gap in materials science between the accuracy and flexibility of first-principles methods and the efficiency of classical force fields. This gap has recently begun to be bridged by the development of machine learning interatomic potentials (MLIPs), which aim to combine near first-principles accuracy with significantly improved computational efficiency.

MLIPs are a rapidly developing class of methods that can significantly outperform classical force fields in terms of accuracy, while remaining orders of magnitude more efficient than first-principles approaches such as DFT\cite{susml}. An MLIP is obtained by training a ML model (most commonly a neural network) on reference data obtained from electronic structure calculations. In this way, the model learns a high-dimensional representation of the potential energy surface (PES) and can subsequently predict energies and forces for previously unseen atomic configurations within the domain of the training data.
A variety of physically motivated MLIP architectures have been recently proposed (\eg MACE\cite{batatiaMACEHigherOrder2023}, NequIP \cite{batznerE3equivariantGraphNeural2022}, GRACE \cite{bochkarevGraphAtomicCluster2024}, So3krates \cite{frankSo3kratesEquivariantAttention2023}, Allegro \cite{musaelianLearningLocalEquivariant2023}, Gaussian regression potential\cite{bartokGaussianApproximationPotentials2010}, and moment tensor potential\cite{shapeevMomentTensorPotentials2016}).
Among these, the MACE family of models has attracted particular attention due to its favourable scaling with the number of chemical elements and strong generalization performance \cite{batatiaMACEHigherOrder2023}. A key feature of MACE is the use of equivariant graph neural networks, which explicitly incorporate rotational symmetries into the model architecture.
MACE models have been successfully employed to study the mechanical and thermal properties of a wide range of defect-free organic and inorganic materials \cite{elenaMachineLearnedPotential2025, kovacsMACEOFFShortRangeTransferable2025, wieserAcceleratingFirstPrinciplesMolecularDynamics2026}.
In contrast, their application to defective materials remains limited \cite{Liu2026,volkmerOntheflyMachineLearning2026}. Addressing this gap is essential for achieving a more accurate characterization of materials properties and for enabling reliable comparisons with experimental measurements \cite{susml}.


Within this context, we assess the robustness of quantum-informed MLIPs for predicting the thermal and mechanical properties of large-scale defective COFs. To this end, we developed a set of MLIPs based on the MACE architecture and trained on a dataset derived from \textit{ab-initio} molecular dynamics of strained carbon-nitride nanosheets \cite{mortazaviExploringStructuralStability2024}.
Our best-performing model, QCOF (``Quantum COF''), combines high computational efficiency with accurate interatomic force prediction and strong transferability to previously unseen chemical environments. Its high inference speed and low memory footprint enable the investigation of low-concentration defects at experimentally relevant scales---a limitation observed for the general-purpose MACE models (MACE-OFF24 and MACE-MPA-0) and their fine-tuned variants considered in this work.
After validating QCOF model suitability, we computed thermal and mechanical properties using non-equilibrium molecular dynamics (NEMD) and uniaxial tensile tests, respectively. A schematic overview of the computational workflow--from database construction, \via MLIP development, to property prediction--is shown in Fig. \ref{fig:01-intro-to-cofs}.
Our results reveal insights into the monolayer response to deformation, showing that even at a low defect concentration, the breaking strength is significantly reduced while the 2D Young's modulus remains practically unaffected.
We also observe a possible relationship between COF stiffness and the degree to which defects influence thermal conductivity.

\section*{Results} 

\subsection*{Benchmarking machine learning interatomic potentials}

An extensive hyperparameter optimization of MLIP models was performed. The models were trained on energies and atomic forces from a diverse dataset comprising over 36,000 conformations of 23 distinct carbon nitride nanosheets, computed using DFT at the PBE+D3 level of theory. The dataset consists of periodic 2D structures containing approximately 100 atoms each. In addition to equilibrium configurations, it includes structures subjected to large tensile strains up to fracture. Further details can be found in the original database publication, see Ref. [\citenum{mortazaviExploringStructuralStability2024}].
Within the MACE architecture, three key hyperparameters were varied: the atomic descriptor dimensionality $D$ (ranging from 32 to 256), the inclusion of angular equivariance $E$ ($E=0$ for invariance, $E=1$ for equivariance), and the interaction cutoff radius $r_{\rm c}$ (ranging from 3 to \SI{7}{\angstrom}). 
The resulting models are labelled according to these hyperparameters, namely, $D-E-r_{\rm c}$.
To enable simulations of large supercells containing defects, computational efficiency was a key consideration. Therefore, the initial benchmark focused on relatively lightweight architectures, ranging from 48-0-4 to 96-1-7 (see Fig. S1 in the Supporting Information (SI)).

To reduce computational cost, the initial screening of MACE models was performed after 50 training epochs.
The effect of $r_{\rm c}$ was assessed using 48-0-$r_{\rm c}$ models with $r_{\rm c}$ ranging from 3 to \SI{7}{\angstrom}. Increasing $r_{\rm c}$ beyond \SI{4}{\angstrom} was found not to yield significant improvements in accuracy. In particular, the variation in force root mean square error (RMSE) across this range is below 10\%, with no systematic decrease as $r_{\rm c}$ increases.
This behaviour can be attributed to the periodic nature of the training configurations. Most structures have unit cell dimensions below \SI{10}{\angstrom}, such that extending the cutoff primarily incorporates redundant information from periodic images and does not substantially improve the description of local chemical environments.
In contrast, variations in model architecture, specifically the atomic descriptor dimensionality $D$ and the inclusion of equivariance $E$, lead to significantly larger differences in accuracy, with force RMSE values spanning approximately \SIrange{40}{80}{\meva} (see Fig. S1 of the SI). Based on these results, a subset of models (48-0-4, 64-0-4, 48-1-4, 64-1-4, 128-1-4, and 128-1-6) was selected for further benchmarking and trained for 500 epochs.
All selected models achieved a validation force RMSE below \SI{60}{\meva}, with the best-performing model (128-1-4) reaching \SI{25.4}{\meva}.

\begin{figure}[t!]
\centering
\includegraphics{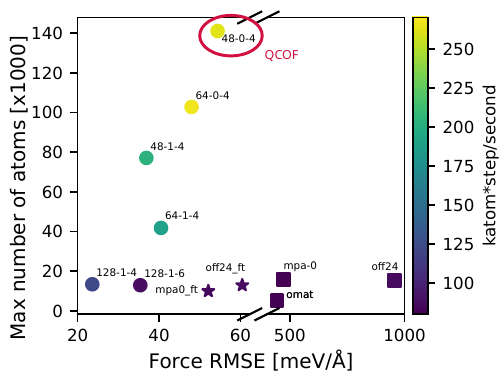}
\caption{Memory efficiency, expressed as the maximum number of atoms that a model can simulate on a single GPU, versus force prediction accuracy on validation data. Model speed is indicated by color, see lateral colorbar. Marker shapes denote the model type: circles represent models trained from scratch on the reference dataset, squares indicate general-purpose MACE models, and stars denote their fine-tuned variants. Inference speed was measured for simulations containing 5,000 atoms. The hyperparameters of the QCOF model lie on the Pareto front with respect to inference speed, memory footprint, and force prediction accuracy.}
\label{fig:02-pareto-opt-mace-models}
\end{figure}

The general-purpose (or “foundational”) models MACE-MPA-0 \cite{batatiaFoundationModelAtomistic2025} and MACE-OFF24 \cite{kovacsMACEOFFShortRangeTransferable2025} were fine-tuned on the same reduced training dataset using the multihead replay method.
MACE-OFF24 was selected as a representative state-of-the-art model for organic systems, while MACE-MPA-0 was included due to its widespread use, broad applicability, and relatively low memory requirements compared to more recent model such as MACE-OMAT.
In addition, the pretrained models without fine-tuning were included as baselines in the benchmark (see Fig. \ref{fig:02-pareto-opt-mace-models}). As expected, these baseline models exhibit the lowest accuracy among the considered approaches, with force RMSE values of approximately \SI{500}{\meva} for MACE-MPA-0 and up to \SI{1000}{\meva} for MACE-OFF24. Their comparatively large architectures also result in slower evaluation speeds and higher memory footprint.
Fine-tuning substantially improves model performance, reducing the force RMSE to the range of \SIrange{50}{60}{\meva}. Nevertheless, these models remain less accurate than the best-performing models trained from scratch on the target dataset.
Based on the trade-off between accuracy, computational efficiency, and memory footprint, the 48-0-4 architecture was found to be the most optimal for further analyses and, from now on, it will be named QCOF model (which stands for ``Quantum Covalent Organic Framework''). 
Indeed, the QCOF model can simulate structures up to 140,000 atoms, which corresponds to an 80 by \SI{100}{nm} monolayer of COF-LZU1 system. The predicted simulation speed for this size is around six computation days on a single NVIDIA H100 GPU per \SI{1}{ns} simulated. 
Such performance makes this model a promising tool both for large-scale simulations, as well as for property screening in high-throughput pipelines.

\begin{table}[t!]
    \centering
    \caption{Benchmark of MACE hyperparameters used to identify the best-performing QCOF model. For this analysis, a reduced dataset containing 10k structures was used, and the models were trained for 500 epochs. In addition to standard root-mean-square errors (RMSE) in force and energy prediction, execution speed and memory efficiency are also reported.}
    \label{table1}
    \begin{tabular}{lllllll}
        \hline\hline
        \multirow{2}{0.5cm}{$E$} &
        \multirow{2}{0.6cm}{$D$} &
        \multirow{2}{1cm}{$r_{\rm c}$ [\AA{}]} &
        \multirow{2}{2.5cm}{kiloatoms/step/\\second} &
        \multirow{2}{2.5cm}{max simulated\\kiloatoms} &
        \multirow{2}{2cm}{force RMSE [eV/\AA{}]} &
        \multirow{2}{2cm}{energy RMSE [meV]} \\
        & & & & & & \\
    \hline
        \multirow{2}{0.5cm}{0}   & 48  & 4 & 261.3 & 141.2 & 52.4 & 1.4 \\
                            & 64  & 4 & 265.7 & 102.8 & 45.3 & 1.1 \\
        \hline
        \multirow{4}{0.5cm}{1}  & 48  & 4 & 201.9 & 77.1  & 35.7 & 1.0 \\
                            & 64  & 4 & 189.7 & 41.8  & 39.4 & 0.7 \\
                            & 128 & 4 & 121.6 & 13.3  & 23.8 & 0.4 \\
                            & 128 & 6 & 86.7  & 12.9  & 33.8 & 0.7 \\
        \hline\hline
    \end{tabular}
\end{table}

\begin{figure}[t!]
\centering
\includegraphics{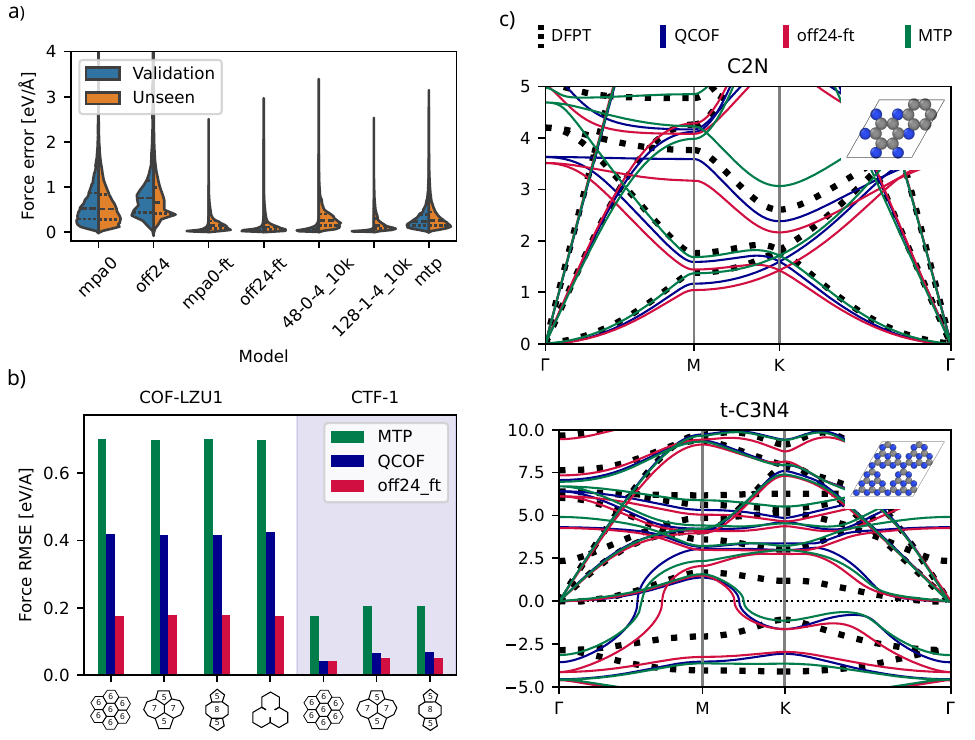}
\caption{Generalisability of different MLIP models. a) Distribution of force errors on validation data and chemical compounds previously withdrawn from training dataset. Points over three standard deviations are omitted for better visibility. The validation dataset contains 500 structures, while the unseen dataset contains 4500. Distributions are rescaled such that maximum width of each is constant. Fine-tuned models show superior generalisability. b) Generalisability of the 48-0-4 architecture to different defects of COF-LZU1 and CTF-1. Both models perform unexpectedly well generalising to unseen defects, retaining almost full accuracy. Structural defects related to pore size seem to not affect the chemical environment perceived by MLIPs. c) Phonon dispersions of C2N and t-C3N4 calculated by VASP DPFT, MTP, fine-tuned MACE-OFF24, and the final COF-MLIP model. Structures of both compounds can be seen on the inset figures (grey -- carbon; blue -- nitrogen).}
\label{fig:03-transferability-and-phonon-benchmarks}
\end{figure}

\begin{table}[t!]
    \centering
    \caption{Validation of the QCOF model for predicting acoustic phonon group velocities, $v_{\rm g}$ (km/s), in pristine COFs. Errors are computed across seven carbon nitride nanosheets, for which $v_{\rm g}$ values were obtained using density functional perturbation theory (DFPT). For comparison, results obtained with the MTP potential, the 128-1-4 model, the fine-tuned MACE-OFF24 model, and the general-purpose MACE-OMAT model are also shown. The full set of $v_{\rm g}$ values is provided in Table S1 of the SI.}
    \label{tab:02-speed-of-sound-errors}
    
\begin{tabular}{llllll}
        \hline\hline
                                  & MTP & QCOF & 128-1-4 & OFF24\_ft & OMAT \\
    \hline
    MAE                             & 0.333 & 0.349 & 0.353 & 0.377 & 0.250 \\
    RMSE                            & 0.542 & 0.416 & 0.399 & 0.418 & 0.307 \\
    $|\Delta v_{\rm g}|_{max}$      & 1.825 & 0.675 & 0.694 & 0.849 & 0.604 \\
    \hline\hline
\end{tabular}
\end{table}

\subsection*{Assessing the generalization of QCOF models}

We now assess the generalization of the QCOF model to unseen chemical compositions. To this end, the model was retrained on a reduced dataset of 10,000 conformations, excluding three of the 23 unique compositions present in the full dataset. For comparison, we also trained a model using the architecture with the lowest force RMSE (128-1-4; see Table \ref{table1}). For clarity, these models are denoted as 48-0-4\_10k and 128-1-4\_10k.
Model performance for energies, forces, and stresses was evaluated and compared to a previously developed Moment Tensor Potential (MTP) trained on the same dataset \cite{mortazaviExploringStructuralStability2024}, as well as to the pretrained (non-fine-tuned) and fine-tuned general-purpose models introduced above (see Fig. \ref{fig:03-transferability-and-phonon-benchmarks}a).
While the fine-tuned models and the 48-0-4\_10k  model trained from scratch exhibit similar accuracy on validation data, their behaviour differs significantly for unseen compositions. The fine-tuned models show substantially better generalization: the force RMSE increases by a factor of four (from 54 to \SI{222}{\meva}) for the 48-0-4\_10k model, compared to a factor of 2.5 (from 60 to \SI{148}{\meva}) for the fine-tuned MACE-OFF24 model.

Next, model transferability to defective structures was evaluated using two COFs: CTF-1 and COF-LZU1 (see Fig. \ref{fig:01-intro-to-cofs}a). The pristine structure of CTF-1 was included in the training data, whereas COF-LZU1 was not seen during training. COF-LZU1 was selected due to its relevance in experimental and theoretical studies \cite{dingConstructionCovalentOrganic2011, guoCatalystderivedHierarchy2D2025, liConstructionFunctionalCovalent2025, majidiFirstprinciplesInvestigationElectronic2025}.
Two main types of localized defects were considered. The first category comprises topological point defects involving 5-, 7-, and/or 8-membered rings, represented here by Stone–Wales (SW) and 5–8–5 defects (see Fig. \ref{fig:01-intro-to-cofs}c). These defects are motivated by the frequent occurrence of non-hexagonal rings in experimentally synthesized COFs \cite{liangOptimalAccelerationVoltage2022}, while maintaining a well-defined and localized structure.
The second category consists of linker or core defects, in which a single Secondary Building Unit is removed, leaving neighbouring functional groups unreacted. For CTF-1, such defects would introduce oxygen atoms, which are not included in the present MLIP parameterization and are therefore excluded. In contrast, oxygen-free core defects can be constructed for COF-LZU1. Together, these chemically motivated defects provide a more realistic representation of experimental disorder than randomly removing individual atoms.

NPT molecular dynamics simulations were performed for five defective structures of both CTF-1 and COF-LZU1, as well as for their pristine counterparts, using the QCOF model at \SI{300}{\kelvin}. A timestep of \SI{0.5}{\femto\second} was used, and each simulation was run for \SI{1}{\nano\second}. 
All simulations remained stable, with no unphysical behaviour observed. Ten equally spaced snapshots were extracted from each trajectory, and atomic forces were recalculated using the QCOF, MTP, and fine-tuned MACE-OFF24 models. Reference forces were obtained from DFT calculations using the same settings as in the training dataset.
For COF-LZU1, a substantial loss in accuracy is observed for all models, with the force RMSE reaching up to \SI{400}{\meva} for the QCOF model. This degradation can be attributed to the presence of chemical environments not represented in the training data. In particular, a significant portion of the error arises from an inaccurate description of the atomic forces associated to the C–N–C angle in the imine linkage (see Fig. S3 in the SI).
Nevertheless, both MACE-based models significantly outperform the MTP model for these defective systems (see Fig. \ref{fig:03-transferability-and-phonon-benchmarks}b and additional results in Fig. S2 of the SI).
For pristine CTF-1, the QCOF and fine-tuned MACE-OFF24 models exhibit comparable accuracy, with force RMSE values of \SI{23}{\meva}. However, the fine-tuned model shows improved transferability to unseen systems, such as COF-LZU1 and defective CTF-1.
Notably, the accuracy for defective structures remains close to that of pristine systems. In CTF-1, a modest decrease in accuracy is observed, with the force RMSE raising to \SI{37}{\meva} for the SW and \SI{39}{\meva} for the 585 defect. In contrast, for COF-LZU1, no significant difference between pristine and defective structures is observed, likely because both represent equally unseen environments for the MLIPs.

An accurate description of phonons is essential for reliable predictions of thermal transport properties \cite{tianImportanceOpticalPhonons2011}. Accordingly, phonon dispersions of selected COF systems were computed using the QCOF potential within the frozen phonon approach (see Fig. \ref{fig:03-transferability-and-phonon-benchmarks}c).
The results are compared with those obtained using the MTP model, the fine-tuned MACE-OFF24 model, and reference data from Density Functional Perturbation Theory (DFPT).
For the C2N system, the phonon dispersion predicted by the QCOF model shows good agreement with DFPT, correctly reproducing dynamical stability and the acoustic branches relevant for thermal conductivity.
In contrast, for the t-C3N4 system, the predictions of all MLIPs are consistent with each other but deviate from the DFPT results. This discrepancy suggests that the training dataset may not adequately sample the relevant atomic configurations required to accurately reproduce the phonon spectrum of this material.
A comparison of phonon dispersions for five additional organic materials is provided in Fig. S4 of the SI. 
We also evaluated the prediction of acoustic phonon group velocities, $v_{\rm g}$, for seven carbon nitride nanosheets (including graphene), as summarized in Table \ref{tab:02-speed-of-sound-errors}. The mean absolute error (MAE) obtained with the QCOF model (0.349 km/s) is comparable to that of the MTP potential, the 128-1-4 model, and the fine-tuned MACE-OFF24 model. However, QCOF achieves lower RMSE (0.416 km/s) and maximum absolute error (0.675 km/s) than both MTP and fine-tuned MACE-OFF24---another compelling evidence of its higher predictive performance. Note that the general-purpose MACE-OMAT model yields the lowest errors in $v_{\rm g}$ prediction; however, its computational efficiency is significantly lower for large-scale simulations (see Fig. \ref{fig:02-pareto-opt-mace-models}). The full set of $v_{\rm g}$ values and boxplot of errors are provided in Table S1 and Fig. S5 of the SI.

\subsection*{Prediction of thermal transport properties}

\subsubsection*{Size effect on thermal conductivity}

It is well established that, despite achieving high accuracy for energies and forces, MLIPs can exhibit significantly larger errors when predicting other material properties \cite{poltavskyCrashTestingMachine2025,susml}.
To assess this, we evaluate the thermal conductivity ($\kappa$) of nitrogenated holey graphene C2N (see structure in Fig. \ref{fig:03-transferability-and-phonon-benchmarks}c), a system for which reliable reference data is readily available \cite{ouyangFirstprinciplesStudyThermal2016, arabhaThermomechanicalPropertiesNitrogenated2021}.
The thermal conductivity was computed using the direct non-equilibrium molecular dynamics (NEMD) approach (see Methods for details). This method is known to be sensitive to system size due to scattering of phonons on thermostat region boundaries \cite{schellingComparisonAtomiclevelSimulation2002}. In this study, we focus on the QCOF model due to its computational efficiency for large-scale simulations (see Fig. \ref{fig:02-pareto-opt-mace-models}). To account for finite-size effects, five simulations with system lengths ranging from 37 to \SI{150}{nm} were performed, and the results were extrapolated to the infinite-length limit. The largest system considered in this analysis contained almost 30,000 atoms.

\begin{figure}[t!]
\centering
\includegraphics{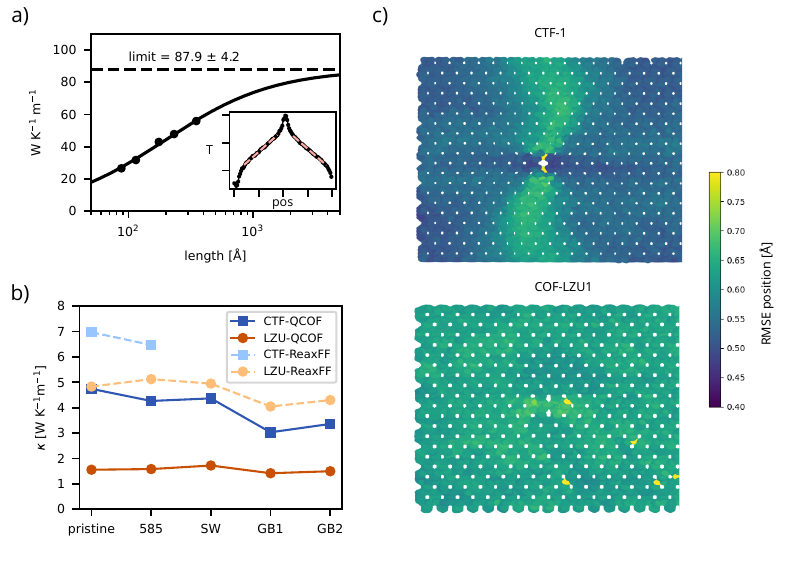}
\caption{a) Model validation to thermal transport prediction. Thermal conductivity of C2N was measured in the zigzag direction at different lengths fitted to Eq. \ref{eq:effective-phonon-mean-free-path}. b) Dependence of thermal conductivity on defects (in order: Stone-Wales, 585 defect, grain boundary type 1, grain boundary type 2) in CTF-1 and COF-LZU1. Missing data for the ReaxFF simulation is caused by simulation crashing due to numerical instabilities. c) Root-mean-square atomic displacement of defective CTF-1 and COF-LZU1 during a heat transport simulation. The influence of the defect (in the middle) is much larger in CTF-1 compared to COF-LZU1.}
\label{fig:04-thermal-conductivity}
\end{figure}

The calculated $\kappa$ values  show a strong dependence on system size, ranging from 23 to \SI{56}{\watt \per \kelvin \per \metre}. After extrapolation, the thermal conductivity converges to \SI{88.9 \pm 9.2}{\watt \per \kelvin \per \metre} and \SI{87.9 \pm 4.2}{\watt \per \kelvin \per \metre} for the zigzag and armchair directions, respectively (see Fig. \ref{fig:04-thermal-conductivity}a for armchair direction and Fig. S7 in the SI for zigzag).
These values are in good agreement with previous computational studies, including \SI{82.22}{\watt \per \kelvin \per \metre} obtained from DFT combined with the Boltzmann transport equation \cite{ouyangFirstprinciplesStudyThermal2016} and \SI{85.5 \pm 3}{\watt \per \kelvin \per \metre} from NEMD simulations using an MTP potential \cite{arabhaThermomechanicalPropertiesNitrogenated2021}. Overall, these results indicate that the QCOF model can reliably reproduce thermal transport properties and is suitable for further large-scale simulations.

\subsubsection*{Thermal transport in defective COFs} \label{sec:defect-thermal-transport}
Thermal conductivities of defective CTF-1 and COF-LZU1 were compared with their pristine counterparts.
For this analysis, we considered a single system size consisting of a simulation box of 34 by 10 rectangular unit cells, which corresponds to 44,880 atoms for pristine COF-LZU1, or 28,560 for pristine CTF-1.
Defects were introduced in the central region of the simulation domain, corresponding to the NVE regions. Specifically, one defect was placed in each NVE region, resulting in a defect concentration of 1/113, or approximately 0.9\% of non-hexagonal pores. For grain boundary (GB) configurations, the heat flow direction was oriented perpendicular to them.

In Fig. \ref{fig:04-thermal-conductivity}b, we present the results obtained using the QCOF model. For comparison, analogous simulations were also performed using the classical ReaxFF force field \cite{vanduinReaxFFReactiveForce2001, chenowethReaxFFReactiveForce2008}. For defective CTF-1, some ReaxFF simulations failed to complete successfully, likely due to numerical instabilities in the force calculations.
Overall, $\kappa$ values obtained with ReaxFF are consistently higher than those predicted by QCOF. This discrepancy may be attributed to the more accurate description of interatomic forces provided by the quantum-informed QCOF model. 
Accordingly, pristine CTF-1 exhibits a $\kappa$ value more than three times higher than that of COF-LZU1 when evaluated with QCOF (see Fig. \ref{fig:04-thermal-conductivity}b). For CTF-1, thermal conductivity decreases with increasing defect size: introducing a point defect reduces $\kappa$ by approximately 10\%, while grain boundaries (GBs) lead to reductions of up to $\sim 30\%$. In contrast, COF-LZU1 does not exhibit a comparable trend, with $\kappa$ remaining nearly unchanged across the considered defect configurations.
This difference can be rationalized by the distinct mechanical stiffness of the two structures. CTF-1 is significantly stiffer than COF-LZU1, making it more sensitive to structural perturbations introduced by defects. Contrarily, the more flexible COF-LZU1 is expected to exhibit shorter phonon mean free paths, such that additional phonon scattering induced by defects has a comparatively smaller impact on thermal transport.
This interpretation is supported by the average atomic displacement shown in Fig. \ref{fig:04-thermal-conductivity}c. A single defect in CTF-1 induces long-range structural perturbations spanning the full width of the system, whereas in COF-LZU1 the defect-induced distortions remain localized.

\subsection*{Exploring elastic properties in large-scale COF}

\begin{figure}[t!]
\centering
\includegraphics{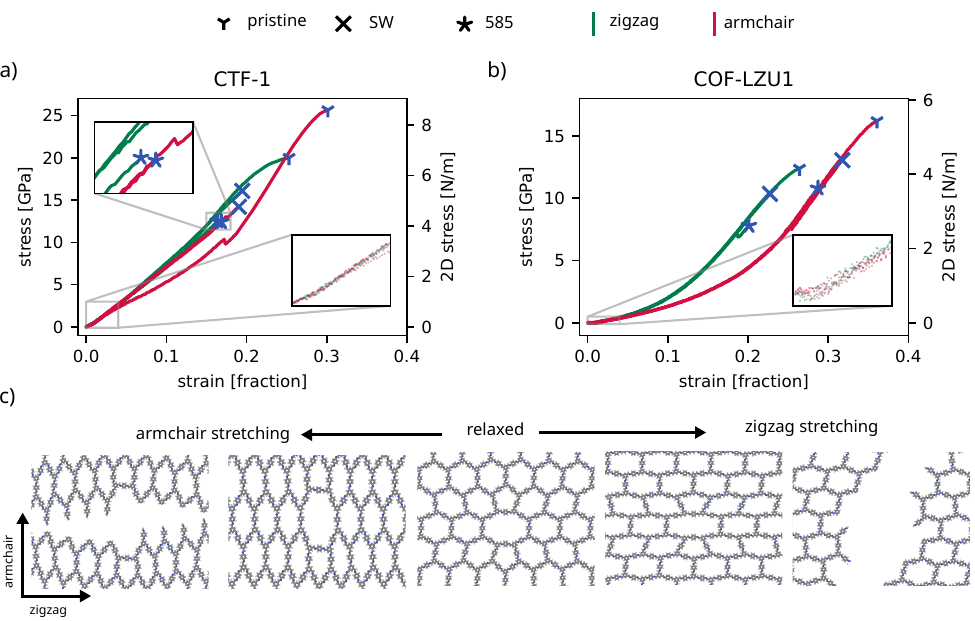}
\caption{Uniaxial tensile strain tests for pristine and defective a) CTF-1 and b) COF-LZU1. Markers indicate the fracture points for each COF configuration. Stress values (in GPa) are calculated assuming a monolayer thickness of \SI{3.35}{\angstrom}. c) Zoomed-in view of COF-LZU1 containing a 585 point defect under tensile strain along the zigzag and armchair directions.}
\label{fig:05-elastic-properties}
\end{figure}

To investigate the effect of defects on the elastic properties of COF monolayers, a series of uniaxial tensile loading simulations were performed using the QCOF model. The defect concentration was kept consistent with that used in the thermal transport simulations, corresponding to one defect per 10 by 17 rectangular supercell.
The resulting system dimensions are approximately 250 by \SI{250}{\nano \metre} for CTF-1 (19,280 atoms) and 380 by \SI{380}{\nano \metre} for COF-LZU1 (22,440 atoms). Note that simulations at this scale would not be feasible using DFT or even more complex MLIPs, such as general-purpose MACE models or their fine-tuned variants.
In Fig. \ref{fig:05-elastic-properties}, we present the stress–strain curves for pristine and defective CTF-1 and COF-LZU1. CTF-1 exhibits a mostly linear response up to very high strain values. The system responds similarly under armchair and zigzag stretching directions, as predicted by linear elasticity theory for materials with hexagonal symmetry \cite{landauTheoryElasticityVolume2012}.
An exception is observed for pristine CTF-1 under armchair stretching, where the QCOF model predicts linker rotation that drives the benzene rings out of the plane. In contrast, COF-LZU1 exhibits anisotropic behaviour at large strains. When stretched in the armchair direction, the stress increases more gradually, allowing the system to withstand higher strain values compared to stretching along the zigzag direction.

The 2D Young's modulus calculated for CTF-1 is \SI{23.2 \pm 0.7}{\newton \per \metre}, which corresponds to \SI{69 \pm 2}{GPa} assuming a single layer thickness of \SI{0.335}{\nano \metre}.
For COF-LZU1, the 2D Young's modulus is \SI{4.9 \pm 0.14}{\newton \per \metre} or \SI{14.6 \pm 0.4}{GPa} for zigzag and \SI{4.11 \pm 0.22}{\newton \per \metre} or \SI{12.3 \pm 0.6}{GPa} for armchair stretching direction. Considering the general shape of the stress-strain curve for COF-LZU1, we can conclude that this difference between two stretching directions is caused by non-linear effects, and the true 2D Young's modulus might be lower than reported.
The simulated fracture stress and Young's modulus of CTF-1 is comparable to results obtained for biaxial strain by Lin \etal using ReaxFF potentials\cite{linTwodimensionalCovalentTriazine2015}. There, the fracture stress of CTF-1 was calculated to be \SI{27}{GPa} and the Young's modulus \qtyrange{60}{66}{GPa},\footnote{Lin \etal quote Young's modulus as \qtyrange{200}{220}{GPa}, but their description matches with another elastic property, the bulk modulus. The Young's modulus of an isotropic material can be calculated as $Y = 2B(1-\sigma)$, where $\sigma$ is the Poisson ratio (quoted by the authors as 0.85).}
which is within 10\% of our results.
The only data found for COF-LZU1 comes from a nanoindentation experiment and is quoted as Young's modulus of \SI{731.4}{MPa}, about 20 times lower than the value calculated here \cite{liConstructionFunctionalCovalent2025}. This significant difference could be caused by the poor crystallinity in experimental samples and significantly less interconnected polymer. In real materials, grain boundaries often involve missing connections, contrary to our simulations where all defects are fully interconnected \cite{duImpactStructuralDefects2024}.
Defects have a significantly higher impact on ultimate tensile strength than on the Young's modulus. Even a single defect in a large 10×17 unit cell lowers the breaking stress by up to 40\%.
The 2D Young's modulus of the studied structures, however, remains virtually unaffected by the presence of defects. This result is consistent with earlier work predicting that the change in 2D Young's modulus caused by defects drops quickly with falling defect concentration, reaching below 1\% for a single defect in a 15 by 15 supercell\cite{bodesheimElasticPropertiesDefective2025}.

The simulations described above were performed at 0 K, \ie they do not account for thermal effects present under experimental conditions. The influence of temperature on elastic properties is twofold.
First, higher temperatures allow the system to explore the potential energy surface more efficiently. This can lead to breaking at lower strain values due to thermal fluctuations, but it may also enable the system to overcome energy barriers and adopt more favourable conformations, potentially relieving stress and delaying fracture.
Second, due to the anharmonicity of chemical bonds, increasing temperature leads to bond elongation and reduced stiffness. Consequently, the Young's modulus decreases at higher temperatures. For example, it has been shown that the Young's modulus of COF-1 and COF-5 decreases significantly at room temperature compared to \SI{0}{K} \cite{wangThermoelasticPropertiesMonolayer2023}.
Additional uniaxial tensile loading simulations at finite temperature for CTF-1 and COF-LZU1 are provided in Fig. S8 of the SI.

\section*{Discussion}

In the present work, we investigated the thermal and elastic properties of defective covalent organic frameworks (COFs) using a quantum-accurate MLIP, namely, QCOF model. Prior to performing these challenging simulations, we systematically assessed several MLIPs in terms of accuracy, computational efficiency, memory footprint, and transferability to unseen chemical environments. 
In addition to analyzing different QCOF model versions, the benchmark included general-purpose models such as MACE-OFF24 and MACE-MPA-0, as well as their fine-tuned variants trained on the dataset employed for developing QCOF model. Based on this evaluation, the QCOF model with a descriptor dimensionality of 48, an invariant representation, and a cutoff radius of 4 \AA{} was identified as the best-performing MLIP across the considered tasks. In particular, it outperformed the fine-tuned general-purpose models in computational efficiency and memory footprint while maintaining comparable accuracy.
For example, the QCOF model enables a \SI{1}{ns} molecular dynamics (MD) simulation of COF-LZU1 containing up to 140,000 atoms with quantum-level accuracy in approximately six days of computation on a single NVIDIA H100 GPU.
The QCOF model is accurate within its chemical domain of expertise characterised by the type of bonding between COF building blocks, and it shows excellent generalisability towards defects. 
Moreover, as shown on the example of COF-LZU1, it allows for stable MD simulations of COFs with unseen chemical linkages at an accuracy still much higher than traditional alternatives (generic ReaxFF parametrisation). 
QCOF predictions were validated on a system well described in literature, C2N, which showed its applicability for elastic property calculations and for NEMD simulations of thermal transport.

Furthermore, the QCOF model generalizes to defective COF structures with minimal loss in force prediction accuracy, thereby enabling large-scale simulations of defect-induced effects on the physical properties of COFs.
In our simulations, CTF-1 showed a clear reduction in thermal conductivity as a response to the introduction of defects, whereas COF-LZU1 remained comparatively insensitive under analogous defect motifs and similar concentrations. 
This contrast suggests that, for COFs, the impact of disorder cannot be inferred from defect density alone. Instead, it appears to depend strongly on the intrinsic mechanical compliance of the framework, the character of its vibrational spectrum, and the extent to which heat transport is already limited by soft, anharmonic, or highly scattered lattice dynamics in the pristine material. Put differently, defects matter most when the pristine framework is stiff enough to support longer phonon mean free paths; when the network is already dynamically disordered and compliant, additional scattering introduced by a localized defect may only weakly perturb the overall heat flow.
This interpretation is consistent with the broader mechanical picture emerging from the tensile simulations. In both COFs, the 2D Young's modulus changed little at low defect concentration, while the ultimate tensile strength dropped much more strongly. That split between stiffness and strength is important. It indicates that small numbers of local defects do not substantially alter the linear-response elasticity of the extended network, but they do act as stress concentrators that determine where failure initiates. For practical applications, this means that a defective COF membrane or thin film may still feel mechanically “stiff” in routine operation, yet remain more vulnerable to catastrophic fracture under large load. In design terms, preserving connectivity and suppressing crack nucleation sites may therefore be more important for reliability than maximizing crystallinity for its own sake.

In summary, we have demonstrated the potential of the optimized QCOF model to perform quantum-informed large-scale simulations of defective COFs and to provide valuable insights into their thermal and elastic properties.
However, the defects and grain boundaries considered here are fully interconnected; that is, no dangling linkers are present. Experimental studies have reported many cases where this assumption does not hold \cite{castanoMappingGrainsBoundaries2021}, particularly for low-angle grain boundaries. The resulting loss of interconnectivity would be expected to produce a more pronounced impact on elastic and thermal properties than observed in the present work.
Flexible COFs exhibit significant thermal fluctuations due to the low frequencies of the phonon modes that restore the structure to its global energy minimum, combined with a rugged potential energy surface containing numerous local minima — some at unit cell geometries far from the room-temperature thermodynamic average. As a result, single unit cell methods that neglect thermal effects are unlikely to capture the complex mechanical behaviour of these materials accurately.
In this context, fast, scalable, and accurate computational approaches, such as the QCOF model presented here, represent an important step forward in the modeling of COFs.
The QCOF model can be further improved by expanding the chemical space of the training dataset. Unlike MTP models, whose complexity scales quadratically with the number of chemical elements, MACE models can more readily accommodate greater chemical diversity. This capability would enable the extension of the approach to oxygen-containing COFs, which are currently beyond the scope of the model.
Another important direction for improvement is the inclusion of van der Waals interactions, which are essential for accurately describing stacked \cite{stacked} and interlocked \cite{interlocked} COFs. In this regard, MLIPs can leverage data-driven approaches to achieve significantly higher accuracy. For example, recent work has demonstrated CCSD(T)-level accuracy for carbon–hydrogen COFs using a delta-learning MLIP built on top of a tight-binding base model to adequately capture long-range electronic interactions \cite{ikedaMachinelearningInteratomicPotentials2026}.
Finally, emerging training strategies such as active learning and model distillation offer efficient routes to expand training datasets at relatively low computational cost. These approaches can further improve the accuracy of future QCOF iterations while preserving their fast inference speed and suitability for large-scale simulations \cite{gardnerDistillationAtomisticFoundation2025,susml}.

\section*{Methods} 

\subsection*{Quantum-accurate machine learning interatomic potential} \label{sec:comp_mlip}

The QCOF models (which stands for ``Quantum Covalent Organic Frameworks'') were developed using the MACE architecture \cite{batatiaMACEHigherOrder2023, kovacsMACEOFFShortRangeTransferable2025}. These models were trained on a dataset of over 36,000 configurations of 23 different two-dimensional nanosheets obtained from \textit{ab initio} molecular dynamics (AIMD) simulations performed with the VASP code at the PBE+D3 level of theory. Additional computational details can be found in Ref. [\citenum{mortazaviExploringStructuralStability2024}].
The first group of models, used for hyperparameter optimisation and generalisability assessment to unseen data, was trained on a reduced dataset consisting of 10,000 randomly selected structures.
Three compounds -- CTF-1, t-C3N4 and C2H5 -- were absent in the reduced dataset. These structures were used to evaluate model transferability.
Additionally, 500 structures were set aside from the reduced dataset as a validation set.
The final QCOF models were trained on the full dataset, with 90\% and 10\% of the data used as training and validation sets, respectively.
For all training runs, a two-stage weighting scheme was employed in the loss function using default weights. During the first 80\% of training epochs, the loss weights were set to 1 for energy, 100 for forces, and 1 for stress. In the second stage, the weights were adjusted to 1000 for energy, 100 for forces, and 10 for stress.

To evaluate the performance of the QCOF models, we additionally considered a Moment Tensor Potential (MTP) trained on the same dataset \cite{mortazaviExploringStructuralStability2024}, as well as general-purpose MACE models MACE-OFF24 and MACE-MPA-0, together with their fine-tuned variants trained on the dataset used to developed the QCOF models. 
Fine-tuning of the general-purpose MACE models was performed using the multi-head replay approach, as described in the official MACE documentation. For each model, a subset of structures containing only carbon, hydrogen, and nitrogen was selected from the original dataset and used as the replay set for the original model head.
For MACE-OFF24, the replay set consisted of 10,000 randomly selected structures. In contrast, for MACE-MPA-0, the original dataset contained only 2,339 structures including these elements; therefore, all of them were used as the replay dataset. The optimiser weights during fine-tuning were kept fixed at 100 for forces and 1 for both energy and stress. Both models were fine-tuned for 100 epochs.

\subsection*{Phonon dispersion}
To validate the accuracy and transferability of developed QCOF models, phonon dispersions were computed using the small displacement method implemented in the Atomic Simulation Environment (ASE) \cite{hjorthlarsenAtomicSimulationEnvironment2017}.
The supercell size chosen was (5,5,1) if the hexagonal unit cell vector was larger than \SI{5}{Å}, and (10,10,0) otherwise. The atom displacement was set to \SI{0.05}{Å}. Phonon group velocities ($v_{\rm g}$) were also calculated to quantify the model performance in predicting acoustic branches. $v_{\rm g}$ is computed as the slope of energy versus wavevector of the two in-plane acoustic phonon branches. The out-of-plane branch was omitted because its theoretical $v_{\rm g}$ ought to be zero.

\subsection*{Thermal transport properties}

In this work, the thermal conductivity ($\kappa$) of pristine and defective COFs was evaluated by running large-scale molecular dynamics (MD) simulations with the developed MLIPs, as implemented in the LAMMPS code \cite{thompsonLAMMPSFlexibleSimulation2022} interfaced with MACE.
 Among the different techniques for computing $\kappa$, we opted for the widely used direct non-equilibrium molecular dynamics (NEMD) method \cite{schellingComparisonAtomiclevelSimulation2002, sandonasEngineeringThermalRectification2015, medranosandonasEnhancementThermalTransport2017}.
In this approach, the simulation cell containing a single layer of the studied COF under periodic boundary conditions was divided into four regions. Two regions (hot and cold reservoirs) were supplied with a constant heat flux using a momentum conserving velocity rescale algorithm \cite{ikeshojiNonequilibriumMolecularDynamics1994}. They were separated by larger NVE regions where the heat could flow freely from the hot to the cold region (see the bottom of panel c) of Fig.~\ref{fig:01-intro-to-cofs}). The length of each thermostatted region was set to 5\% of the total simulation box length.
Before the heating and cooling were established, the full system was equilibrated under NPT conditions at zero pressure and a temperature of 300K. No temperature drift was observed during the simulations even after the thermostat was removed. The simulation time step was set to 0.5 femtoseconds for all MLIPs and to 0.25 femtoseconds for ReaxFF to improve convergence.

Then, $\kappa$ was calculated based on Fourier's law:
\begin{equation}
J_\mu = - \sum_\nu \kappa_{\mu \nu} \frac{\partial T}{\partial x_\nu} \ ,
\end{equation}
which relates the imposed heat flux to the measured temperature gradient. 
Because the temperature gradients oscillate over time and the temperature values are strongly time-correlated, with typical correlation times between 20 and \SI{100}{\pico\second} (see Fig.~S6 in the SI), the uncertainty of the calculated temperature gradients was estimated by standard error:
\begin{equation}
    \text{SE} = \frac{\sigma}{\sqrt{n_\text{eff}}} \ ,
\end{equation}
where $n_\text{eff} = t_\text{total}/\tau_\text{autocorrelation}$ stands for the effective number of independent samples.
For comparison, we also present results of $\kappa$ obtained with a ReaxFF parametrisation developed for oxidation of hydrocarbons \cite{chenowethReaxFFReactiveForce2008}.

The non-equilibrium setup defined in such a way may induce intense phonon scattering at the region boundaries, which in turn reduces the observed heat flux. Consequently, $\kappa$ calculated for a finite COF fragment is lower than its infinite-length limit. This finite-size effect can be modelled by considering the effective phonon mean free path, given by
\begin{equation}
    \frac{1}{l_\text{eff}} = \frac{1}{l_\infty} + \frac{4}{L_x},
    \label{eq:effective-phonon-mean-free-path}
\end{equation}
where $L_x$ is the total length of the simulation cell \cite{schellingComparisonAtomiclevelSimulation2002}. Since thermal conductivity $\kappa$ in kinetic transport theory is directly proportional to the phonon mean free path, a linear extrapolation of $1/\kappa$ as a function of $1/L_x$ can be performed. The infinite-length thermal conductivity is then obtained by extrapolating to the limit $1/L_x = 0$. This extrapolation procedure was applied only for benchmarking the QCOF model in the calculation of $\kappa$ for C2N.

\subsection*{Mechanical properties}

Stress–strain curves were obtained from uniaxial tensile loading simulations at zero temperature. A 10 × 17 supercell (approximately square in shape) of the selected COF, with periodic boundary conditions and optionally containing a single central defect, was statically optimized to its minimum-energy configuration.
The simulation cell was then incrementally stretched in steps of 0.1\% along the $x$ or $y$ direction. After each increment, atomic positions were relaxed using the conjugate gradient algorithm. The strained direction was held fixed, while the simulation box was allowed to relax to zero pressure in the direction perpendicular to the applied strain.
2D Young's modulus (Y) was extracted from a linear fit of stress versus strain:
\begin{equation*}
Y = \frac{\Delta S}{\Delta \epsilon}
\end{equation*}
in the strain range from 0 to 5\% for CTF-1 and from 1\% to 5\% for COF-LZU1. Data between 0 and 1\% was discarded due to high non-linearity.

\section*{Data availability}
The dataset containing the structures of covalent organic frameworks used for training the MLIP models is publicly available at \href{https://data.mendeley.com/datasets/jrhww36ccv/1}{https://data.mendeley.com/datasets/jrhww36ccv/1}.
The models, codes, and additional benchmarking datasets used in this work can be found in \href{https://github.com/Alo-koder/QCOF}{Repo-QCOF}.


\bibliography{cofml}

\section*{Acknowledgements} 
A.S. and G.C. acknowledge funding by the DFG project "Data-Driven Characterization of (A)Chiral 2D Polymers" (CRC1415 - C04)
L.M.S and G.C. gratefully acknowledge the funding by the German Research Foundation (DFG) under the Cluster of Excellence CeTI: Centre for Tactile Internet with Human-in-the-Loop (EXC 2050/2, Project ID 390696704) and Cluster of Excellence CARE: Climate-Neutral And Resource-Efficient Construction (EXC 3115, Project ID 533767731). 
Funding by the DFG via the ``Responsible Electronics in the Climate Change Era – REC²'' Cluster of Excellence (EXC 3035, Project-ID 533607596) is gratefully acknowledged.
We thank the Center for Information Services and High-Performance Computing (ZIH) at TU Dresden for providing the computational resources and technical support.

\section*{Author contributions}

The work was initially conceived by AS and LMS, and its design was developed with input from DB, BM, and GC. AS and LMS trained the MLIPs using MACE, and conducted performance analyses. AS also evaluated the capabilities of MLIPs to compute the thermal and elastic properties of large-scale COFs. AS, LMS, and DB drafted the original manuscript. All authors discussed the results and contributed to the final version of the manuscript.

\section*{Competing interests}

The authors declare no competing interests.

\end{document}


 \fancyhead{}
 \renewcommand{\headrulewidth}{1pt}
 \renewcommand{\footrulewidth}{1pt}
 \setlength{\arrayrulewidth}{1pt}
 \setlength{\columnsep}{6.5mm}
 \renewcommand{\figurename}{Fig.~S\!\!}
 \renewcommand{\tablename}{Table~S\!\!}

 \begin{center}
 \noindent\LARGE{Supplementary Information (SI) for:}
 \vspace{0.3cm}

 \noindent\LARGE{\textbf{Data-Driven Thermal and Mechanical Modeling of Defective Covalent Organic Frameworks}}
 \vspace{0.6cm}

 \noindent\large{\textbf{Aleksander Szewczyk\textit{$^{a}$}, Leonardo Medrano Sandonas\textit{$^{a}$}$^{\ast}$, David Bodesheim\textit{$^{a}$}, Bohayra Mortazavi\textit{$^{b,c}$}, and Gianaurelio Cuniberti\textit{$^{a,d,e,f}$}$^{\ast}$}} \vspace{0.5cm}

 \noindent{\textit{$^{a}$~Institute for Materials Science and Max Bergmann Center of Biomaterials, TUD Dresden University of Technology, 01062 Dresden, Germany.}}\\[0.6em]
  \noindent{\textit{$^{b}$~Institute of Photonics, Department of Mathematics and Physics, Leibniz Universität Hannover, Welfengarten 1A, 30167, Hannover, Germany.}}\\[0.6em]
    \noindent{\textit{$^{c}$~Cluster of Excellence PhoenixD, Leibniz Universität Hannover, Welfengarten 1A, Hannover 30167, Germany.}}\\[0.6em]
      \noindent{\textit{$^{d}$~Dresden Center for Computational Materials Science (DCMS), TUD Dresden University of Technology, 01062 Dresden, Germany.}}\\[0.6em]
        \noindent{\textit{$^{e}$~Cluster of Excellence CARE, TU Dresden and RWTH Aachen, Germany.}}\\[0.6em]
   \noindent{\textit{$^{f}$~Cluster of Excellence CeTI, TU Dresden, Germany.}} \\[1em]
  \noindent{$^{\ast}$~Corresponding authors: Leonardo Medrano Sandonas (\texttt{leonardo.medrano@tu-dresden.de}) and Gianaurelio Cuniberti (\texttt{gianaurelio.cuniberti@tu-dresden.de})}\\[0.6em]
 \end{center}
 
\vspace{0.5in}
\clearpage 

\section{Benchmarking QCOF model}

\begin{figure}[hbtp]
    \centering
    \includegraphics[width=0.6\linewidth]{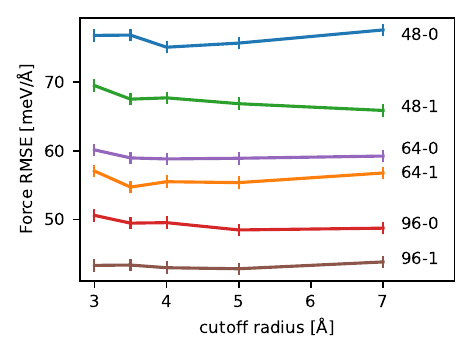}
    \caption{Model accuracy vs cut-off distance for variants of the QCOF model trained for 50 epochs. Increasing the cut-off radius does not bring meaningful accuracy improvements.}
    \label{fig:placeholder}
\end{figure}

\begin{figure}[hbtp]
    \centering
    \includegraphics[width=0.7\linewidth]{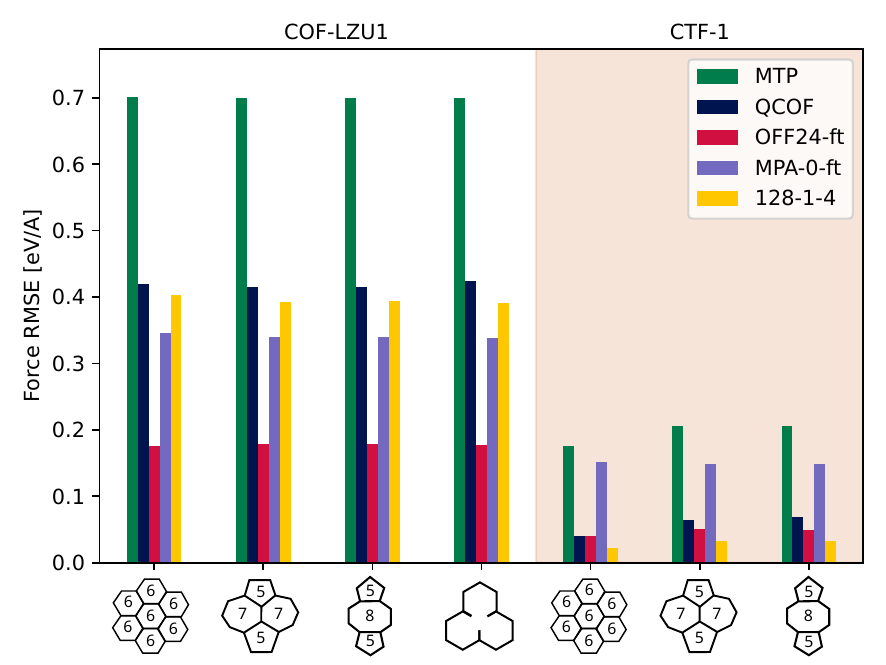}
    \caption{Generalisability of fine-tuned models to unseen defects on a force benchmark. 'ft' signifies fine-tuned models. The model based on MACE-OFF24, despite slightly worse accuracy in validation dataset, is able to generalise to COF-LZU1 much better than MACE-MPA0 due to having seen more similar organic structures in its original training dataset.}
    \label{fig:placeholder}
\end{figure}

\begin{figure}[hbtp]
    \centering
    \includegraphics[width=0.6\linewidth]{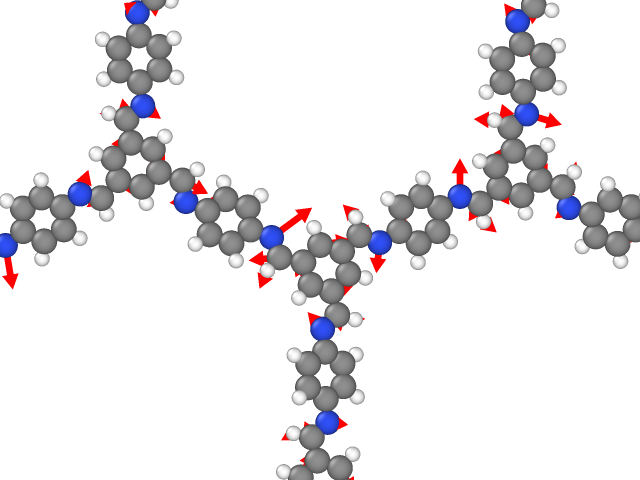}
    \caption{Local force errors of the QCOF model compared to reference DFT calculations of COF-LZU1. The arrow length is proportional to force error. Majority of the error can be attributed to the incorrect equilibrium size of the C-N-C angles.}
    \label{fig:placeholder}
\end{figure}

\newpage

\section{Phonon calculations}

\begin{figure}[hbtp]
    \centering
    \includegraphics[width=0.8\linewidth]{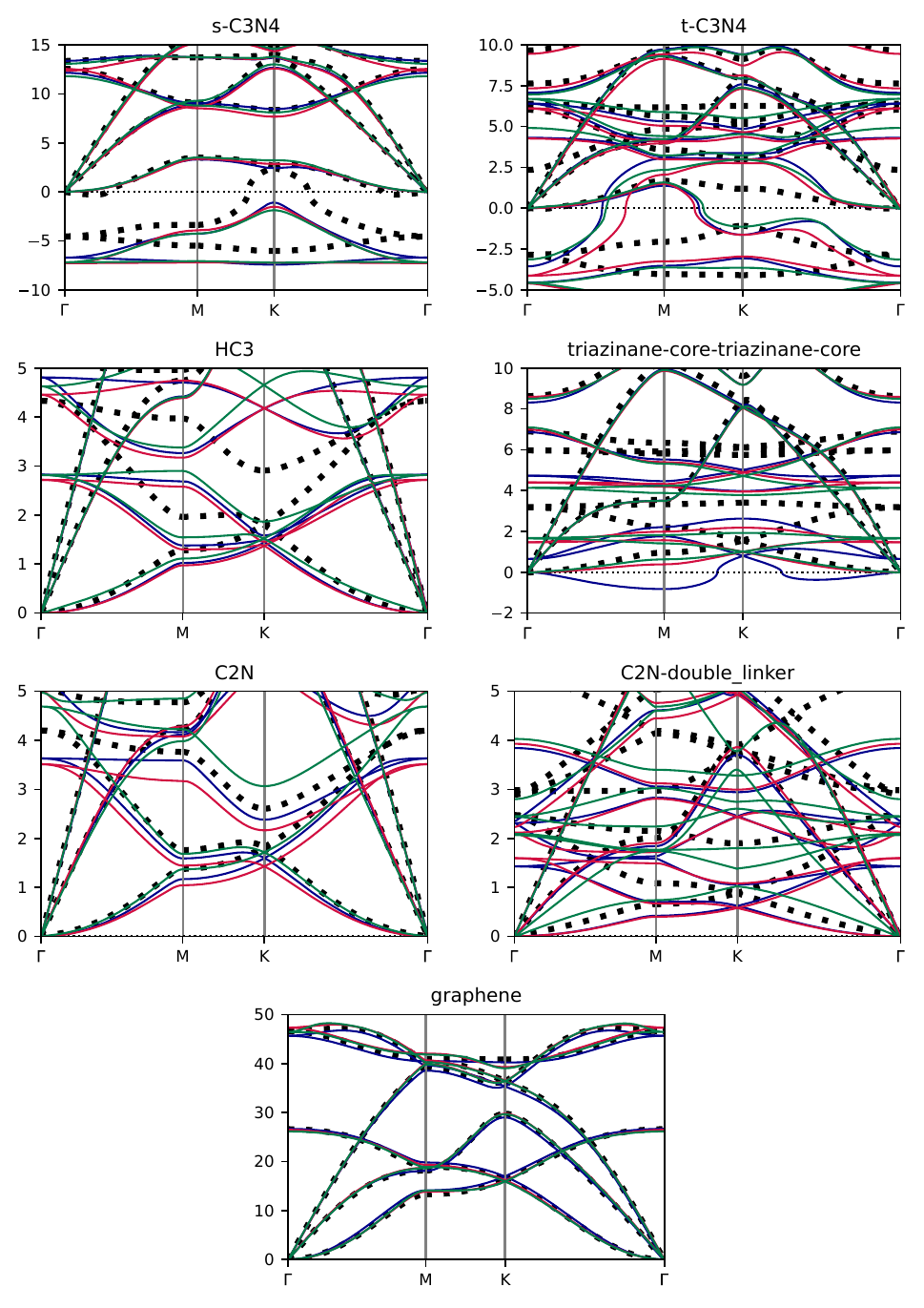}
    \caption{Phonon dispersions calculated with DFPT (black dotted line), the QCOF model (blue), fine-tuned MACE-OFF24 (red) and the MTP potential (green). $y$ axis: phonon frequency in THz.}
    \label{fig:placeholder}
\end{figure}

\begin{figure}[hbtp]
    \centering
    \includegraphics[width=0.7\linewidth]{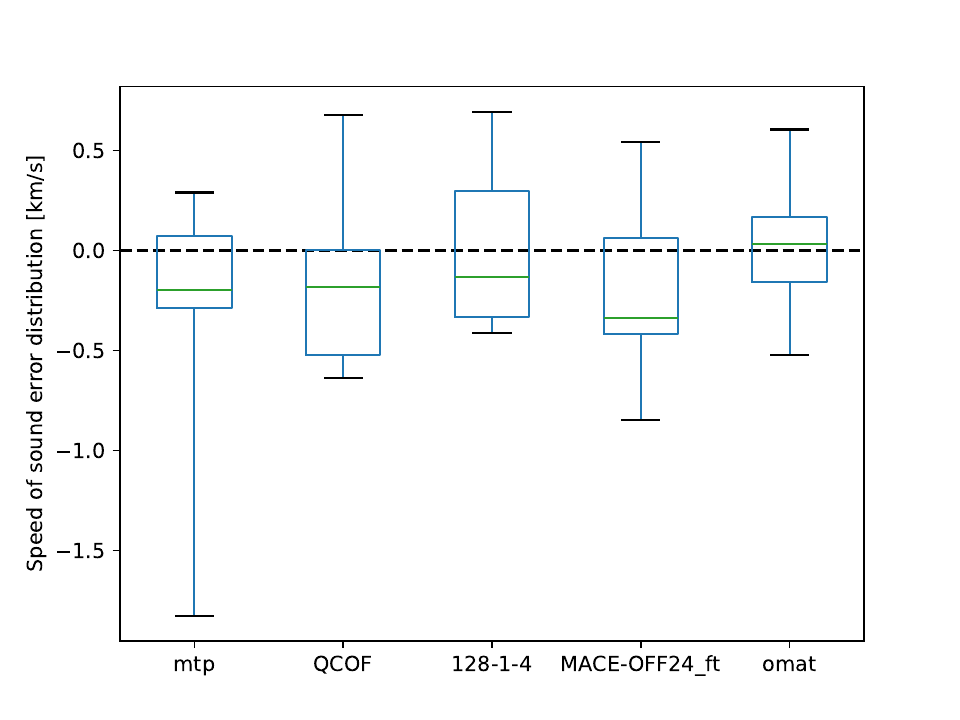}
    \caption{Speed of sound error distribution calculated on 7 structures listed in the figure above.}
    \label{fig:placeholder}
\end{figure}

\begin{table}[hbtp]
    \caption{Calculated phonon group velocities [km/s] and their errors with DFPT calculation as reference. OFF24\_ft signifies the fine-tuned MACE-OFF24 model. For each structure, two in-plane acoustic phonon branches are considered.}
    \centering
    
\begin{tabular}{l|r|r|r|r|r|r}
    \hline
    system & DFPT & MTP & QCOF & 128-1-4 & OFF24\_ft & OMAT \\
    \hline
    graphene                            & 13.5 & 13.6 & 14.0 & 14.2 & 14.1 & 13.4 \\
                                        & 9.4  & 9.2  & 9.4  & 9.6  & 9.6  & 9.3  \\
    C2N                                 & 10.3 & 10.1 & 10.1 & 10.2 & 10.1 & 10.4 \\
                                        & 9.9  & 9.7  & 9.7  & 9.6  & 9.6  & 9.7  \\
    HC3                                 & 7.5  & 5.7  & 6.9  & 7.2  & 7.2  & 7.0  \\
                                        & 11.1 & 11.2 & 10.9 & 11.4 & 10.7 & 11.0 \\
    triazinane-core-triazinane-core     & 9.4  & 9.7  & 10.1 & 10.1 & 9.5  & 10.0 \\
                                        & 21.4 & 21.6 & 21.4 & 22.0 & 21.6 & 21.6 \\
    C2N double linker                   & 16.3 & 15.9 & 15.7 & 15.9 & 16.0 & 16.2 \\
                                        & 17.2 & 16.7 & 16.6 & 16.8 & 16.7 & 17.3 \\
    s-C3N4                              & 15.7 & 15.4 & 15.5 & 15.3 & 15.2 & 15.4 \\
                                        & 14.2 & 13.9 & 13.6 & 14.0 & 13.8 & 14.3 \\
    t-C3N4                              & 17.9 & 17.8 & 17.5 & 17.7 & 17.0 & 18.0 \\
                                        & 16.1 & 16.2 & 16.3 & 16.4 & 15.6 & 16.7 \\
    \hline
    Root Mean Square Error              & N/A & 0.333 & 0.349 & 0.353 & 0.377 & 0.250 \\
    Mean Absolute Error                 & N/A & 0.542 & 0.416 & 0.399 & 0.418 & 0.307 \\
    Maximum Absolute Error              & N/A & 1.825 & 0.675 & 0.694 & 0.849 & 0.604 \\
    \hline
\end{tabular}
\end{table}

\clearpage

\newpage

\section{Thermal conductivity}

\begin{figure}[hbtp]
    \centering
    \includegraphics[width=0.85\linewidth]{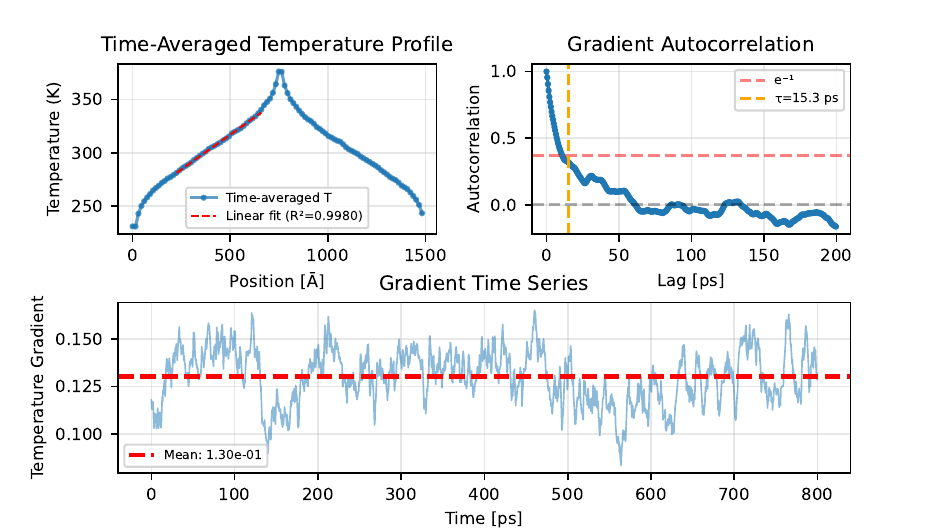}
    \caption{The statistical analysis of the thermal profile timeseries for \SI{150}{\nano \metre} long monolayer of C2N. The thermal gradient shows a strong autocorrelation in time, which necessitates a careful approach to uncertainty estimation.}
    \label{fig:placeholder}
\end{figure}

\begin{figure}[b!]
    \centering
    \includegraphics[width=0.5\linewidth]{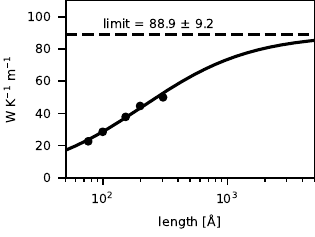}
    \caption{Thermal conductivity of C2N in the zigzag direction extrapolated to infinite system size (\cf Figure 4a in main text).}
    \label{fig:placeholder}
\end{figure}

\clearpage

\section{Elastic properties}

\begin{figure}[hbtp]
    \centering
    \includegraphics{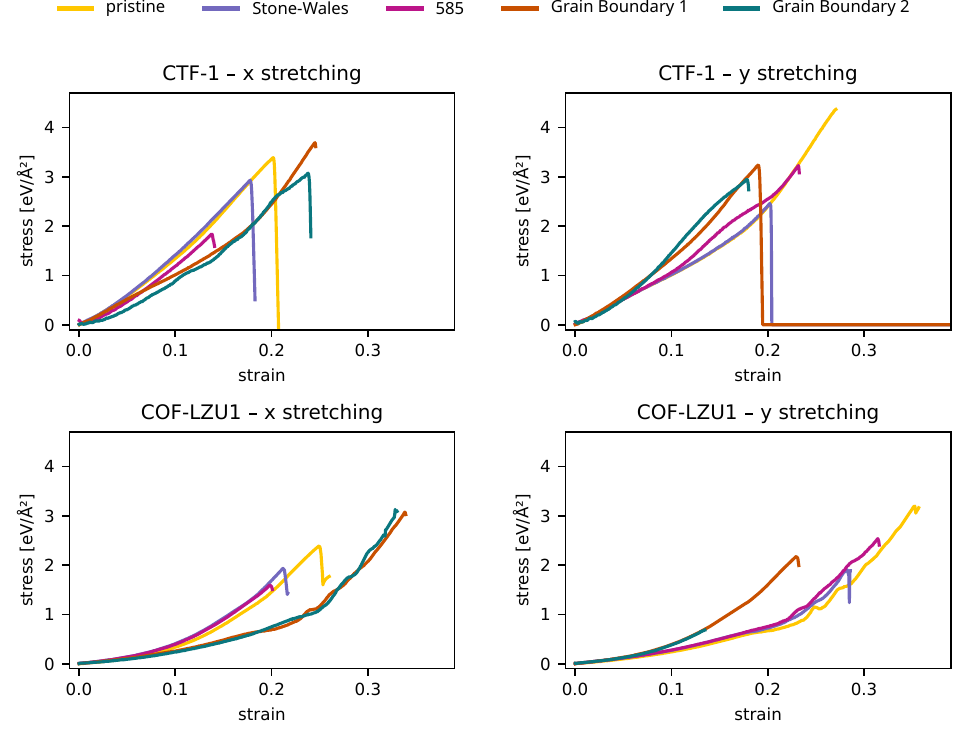}
    \caption{Full stress-strain curves of uniaxial tensile loading simulations at finite temperature (\SI{300}{\kelvin}). The curves break abruptly when the simulation crashes due to numerical instabilities caused by a sudden release of accumulated stress. These simulations are sensitive to initial conditions -- the oscillations in stress are expected to disappear with a bigger statistical sample of simulation runs.}
    \label{fig:placeholder}
\end{figure}